\documentclass[letterpaper, 10 pt, conference]{ieeeconf}  % Comment this line out if you need a4paper

\IEEEoverridecommandlockouts                              % This command is only needed if 
\usepackage{amsmath}
\usepackage{graphicx}

\usepackage{mathpazo}
\usepackage{subcaption}
\usepackage{multirow}
\usepackage{authblk}

\title{\LARGE \bf
Brain-Computer Interface in Virtual Reality
}

\author{Reza Abbasi-Asl$^{1}$, Mohammad Keshavarzi$^{2}$, and Dorian Yao Chan$^{3}$% <-this % stops a space
\thanks{$^{1}$Reza Abbasi-Asl is with Department of Electrical Engineering and Computer Sciences, University of California, Berkeley, CA, USA. {\tt\small abbasi@berkeley.edu}}%
\thanks{$^{2}$Mohammad Keshavarzi is with Department of Architecture, University of California, Berkeley, CA, USA. {\tt\small  mkeshavarzi@berkeley.edu}}
\thanks{$^{3}$Dorian Yao Chan is with Department of Electrical Engineering and Computer Sciences, University of California, Berkeley, CA, USA. {\tt\small dorian8x8@berkeley.edu}}
}

\begin{document}

\maketitle
\thispagestyle{empty}
\pagestyle{empty}

%%%%%%%%%%%%%%%%%%%%%%%%%%%%%%%%%%%%%%%%%%%%%%%%%%%%%%%%%%%%%%%%%%%%%%%%%%%%%%%%

\begin{abstract}

We study the performance of brain computer interface (BCI) system in a virtual reality (VR) environment and compare it to 2D regular displays. First, we design a headset that consists of three components: a wearable electroencephalography (EEG) device, a VR headset and an interface. Recordings of brain and behavior from human subjects, performing a wide variety of tasks using our device are collected. The tasks consist of object rotation or scaling in VR using either mental commands or facial expression (smile and eyebrow movement). Subjects are asked to repeat similar tasks on regular 2D monitor screens. The performance in 3-D virtual reality environment is considerably higher compared to the to the 2D screen. Particularly, the median number of success rate across trials for VR setting is double of that for the 2D setting (8 successful command in VR setting compared to 4 successful command in 2D screen in 1 minute trials). Our results suggest that the design of future BCI systems can remarkably benefit from the VR setting.

\end{abstract}

\section{Introduction}

The development of brain-computer interface (BCI) systems have valuable applications in areas such as medicine \cite{Pinheiro2016Wheelchair,Pfurtscheller1993Brain}, robotics \cite{Berger2008Brain,Bell2008Control}, and human entertainment industry \cite{Alomari2014EEG,Krepki2007Berlin}. Helping people with movement disabilities to retract their motor functionalities through mental commands and communicate with the digital world is one of the most important consequences of such technology. A BCI system often consists of three main components. First, a module to record the activity of neurons in the brain (either single neuron \cite{Maynard1997Utah} or average of thousands of neurons \cite{Pfurtscheller2007EEG}). Second, a digital or mechanical environment (such as computers or robotic arms) that the user intend to control. Third, an interface that processes the brain signal and translate it to an actionable command in the target environment. 
%Wide neuroprosthetics applications that aim at restoring damaged hearing, sight and movement.

While there have been major progress in designing and employing each of BCI three components over the last decades, some limitations in each category are yet to be addressed. The available devices that record from the brain often have a very low signal-to-noise ratio \cite{Teplan2002Fundamentals}. Researchers have shown a remarkable performance for the invasive BCI systems that receive recordings from single neurons \cite{Maynard1997Utah}. However, developing BCI system using non-invasive devices such as Electroencephalography (EEG) \cite{Teplan2002Fundamentals} is a much more challenging task. EEG devices are less expensive and more user-friendly compared to invasive recording systems, but the measures signal using each electrode in EEG is the average activity of thousands of neurons. Additionally, because EEG electrodes are placed along the scalp, EEG signals are often noisy and distorted \cite{Teplan2002Fundamentals}. Therefore, it is difficult to recover the underlying activity of the brain from EEG recordings. Another limitation of the EEG device is the number of channels. A higher signal to noise ratio can be achieved using an EEG with large number of electrodes. However, these high-bandwidth EEG devices are often less user-friendly compared and have higher cost. In this paper, we are interested in applications of BCI that is adaptable to daily life of users interacting with digital devices. Therefore, we accept the challenge of the brain recordings with low signal-to-noise ratio and limit our study to the user-friendly wearable EEG headsets with small number of electrodes.

EEG devices often record not only the average activity of the brain areas, but also the facial expressions such as eyebrow and lip movement \cite{Badcock2013Validation}. To decode the target command, EEG-enabled BCI device primarily benefit from state-of-the-art machine learning algorithms. Methods such as deep neural networks \cite{Zhang2017Converting,Carvalho2017Deep}, generative models \cite{Palazzo2017Generative} and Bayesian models \cite{Phillips2005empirical} have shown satisfactory performance in these systems.

Being able to efficiently interact with a digital device is a necessity in many of today's real-life applications. These applications range from simple tasks such as using cellphones or moving a cursor on a screen to much more complicated tasks such as controlling robotic arm movements. In all of these applications, high bandwidth in communication is considered one of the most important aspects of the human-machine interactions \cite{Cheng2002Design}. In this study, we are interested in the limitations of user interaction with a virtual reality (VR) environment and seek BCI solutions to increase the bandwidth. VR is particularly an important tool in fields such as design \cite{Sherman2002Understanding}, education \cite{Sanchez2000Design} and communication \cite{Biocca2013Communication}. To increase the communication bandwidth between user and VR environment, we study the application of EEG-based BCI systems in VR. With the recent progress in development of virtual and augmented reality, it has been necessary to analyze, study, and evaluate the performance of BCI devices in controlling these virtual environments. Being able to use communication channels that do not need hand movement and gestures is essential in this setting. There has been a limited number of studies trying to build such a connection. Additionally, several industrial start-ups such as Neurable \cite{neurable} are aiming to improve the reliability of BCI-driven VR. 

In this paper, we design and implement protocols to quantitatively study the possibility of directly controlling the virtual reality environment using commands translated from brain activity. Additionally, we compare the performance of BCI in 3D VR environment to the 2D regular screen. The rest of the paper is organized as follows. In section 2, we introduce our set up to control a virtual reality application using an EEG device. The experimental protocols followed by our main results are presented in section 3. We conclude and discuss our future directions in section 4. 

%Our results show that BCI in VR is more accurate because the users are more focused in a virtual 3D environment.

%Compare Mental and Facial commands in 2D screens and VR displays and Applications to practical gameplay

\section{Device Setup}

\begin{figure}
\centering
\includegraphics[width=0.48\textwidth]{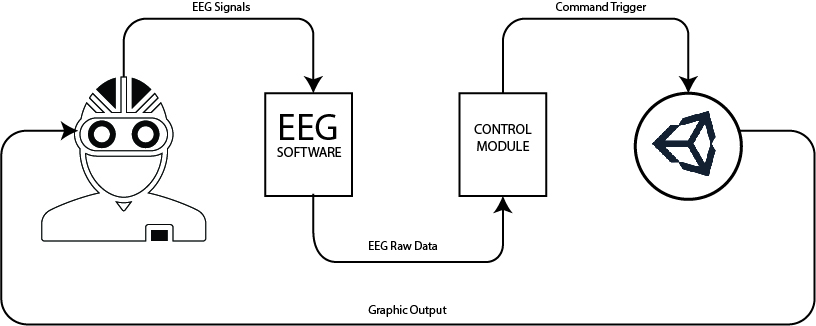}
\caption{Flowchart of the pipeline}
\label{fig:flow}
\end{figure}

Figure \ref{fig:flow} shows the flowchart of our BCI-enabled VR. Our pipeline consists of an EEG module, a VR module and and interface. Our specific choice for each module is summarized in this section.

\subsection{EEG}
  
To record the brain activity, we use Emotiv EPOC®+ EEG headset. This EEG device has 14 electrodes with saline based wet sensors. These electrodes do not require any gel and therefore is a better match to our application compared to regular wet electrodes. 14 channels is sufficient in our application and has enough bandwidth to enable us to control VR/AR device. The user is able to wear this EEG headset together with the AR/VR headset. Therefore, this design is user0friendly. We did not chose Electro-cap EEG because it gradually squeezes the user’s head and becomes uncomfortable for the user after a short time.

\begin{figure*}
\centering
\includegraphics[width=1\textwidth]{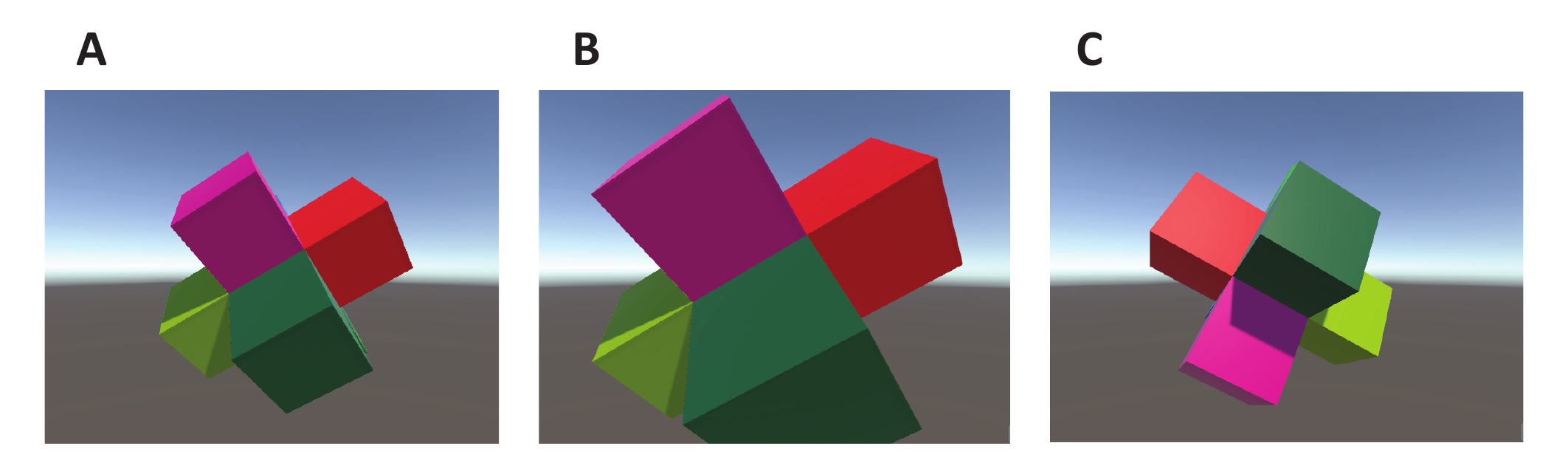}
\caption{A. 3D cross neutral. B. 3D cross scaled. C. 3D cross rotated. }
\label{fig:shapes}
\end{figure*}

\subsection{Virtual reality module}

To train and execute mental and facial expressions, we use the Unity3D game engine to visualize the outcome of the user’s commands in the 2D screen and virtual reality. In this application, 7 boxes shaping a 3D cross each with different colors assigned are placed in the middle of the users view (Figure \ref{fig:shapes}.A). The background and lighting of the scene is designed to be simple as possible to avoid user distraction. In the event of a specific command, the 3D cross can move forward and backward, rotate, scale and change color (Figures \ref{fig:shapes}.B and \ref{fig:shapes}.C) . In the training process, such transforms are initiated from the start of the training session and last 10 seconds while in the execution process the transforms occur when the mental command are triggered and turn off when neutral conditions are detected. In such events, the occurrence of the transformation gradually dissolves to minimize distraction of the inconsistency of the mental and facial commands. The virtual reality device used in this experiment is the HTC Vive, as the larger size of the headset – comparing to the Oculus Rift - allows easier allocation of EEG sensors.

\begin{figure}
\centering
\includegraphics[width=0.48\textwidth]{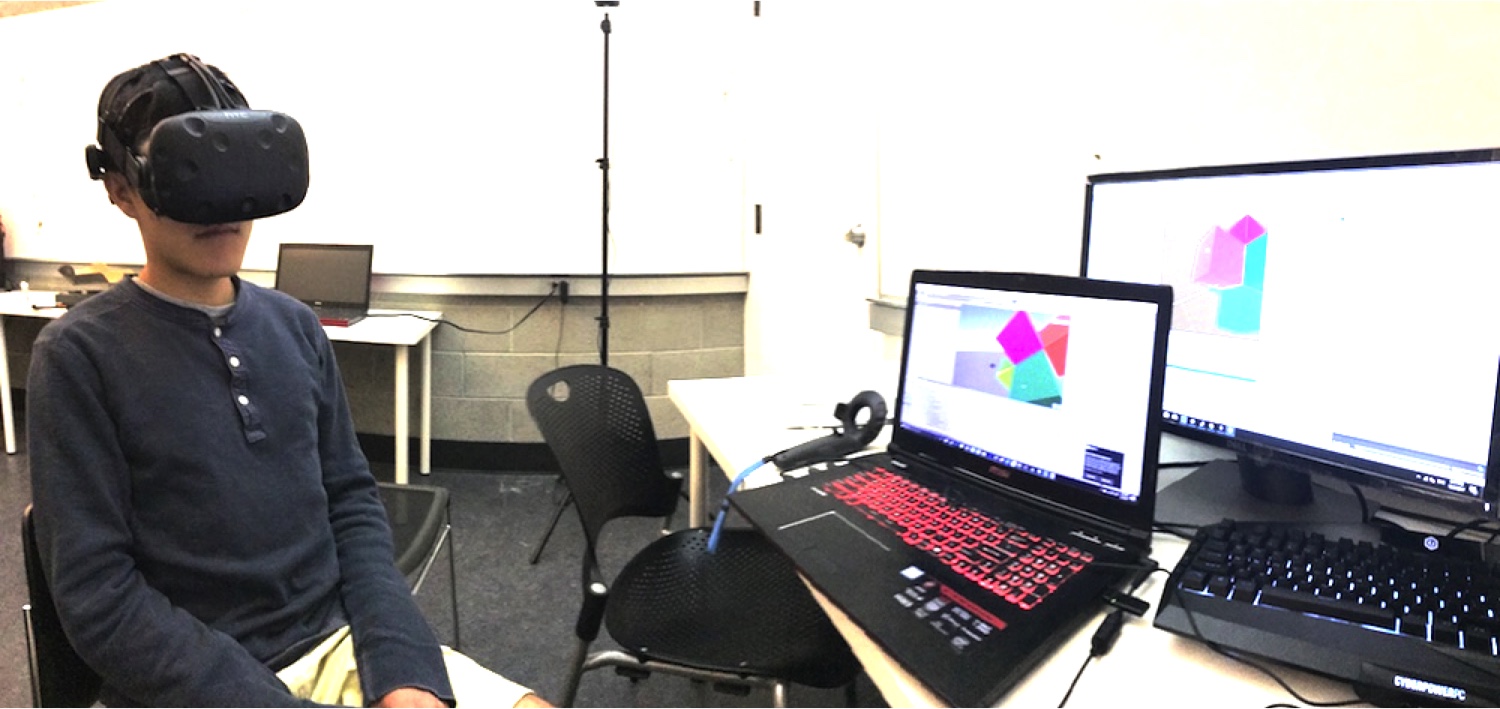}
\caption{Setup of the experiment}
\label{fig:setup}
\end{figure}

\subsection{Interface}

We use Emotive mental-command software package to train the model. Based on the notes from developers \cite{neurable}, this software includes the following modules to process the EEG signals. First, a basic filtering and real-time classification to knock out spikes and other non-biological signals/noise. In this step, a certain amount of ‘muscle noise’ (below a threshold) is allowed to give a good experience for novice users and assist the learning process.
The training procedure is built upon EEG features such as frequency content and spatial distribution of components.
Correlation analysis technique is used to reduce the input dimensionality to a manageable level for the amount of training data available. Final classification is done by calculating the relative likelihood of a given observation belonging to each of the trained classes, assigning the point to the command with the highest posterior probability.

\section{Results}

\subsection{Data collection}

To quantitatively compare the performance of BCI system between a 2D display and 3-D virtual reality environment, we designed the following protocol to collect data from subjects. We first trained commands by having the user repeatedly attempt a command over 5 minutes. We then performed 1 minute trials, where we asked the participant to attempt a particular command every 6 seconds (10 commands per trial). We repeated this process for 10 times (100 commands for each condition). We recorded the success rate, as well as the number of false positives. Participants attempted to use commands to rotate or push objects. On some trials, we also asked the participants to alternate between them, in order to compare the accuracies of different command techniques. Figure \ref{fig:setup} shows a picture of our experiment setup.

\subsection{Analysis}

\begin{figure*}
\centering
\includegraphics[width=1\textwidth]{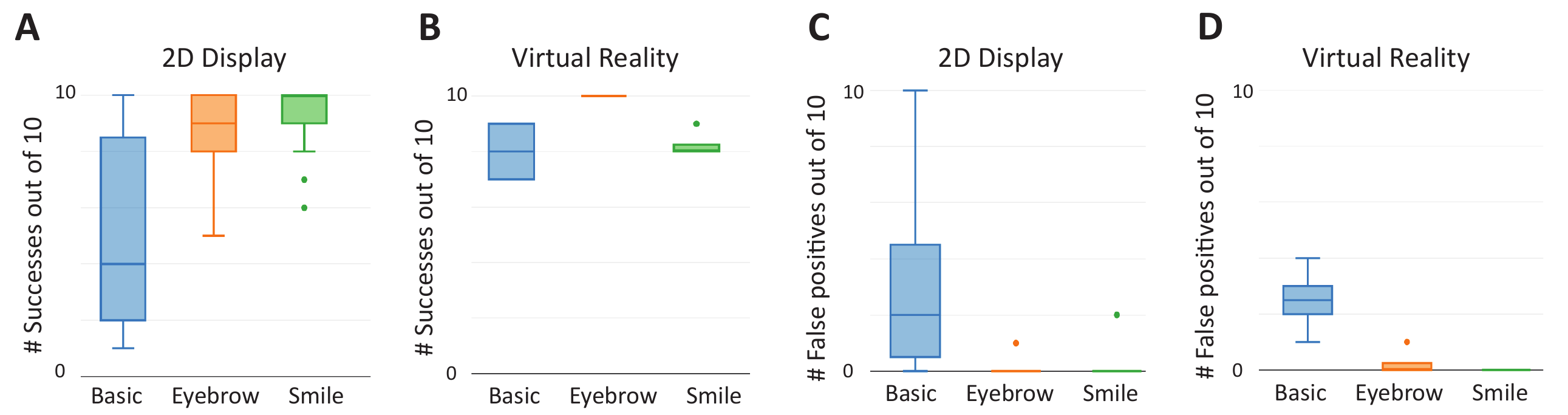}
\caption{Number of successes for binary experiment in 2D display (A) and VR setting (B). Number of false positives for binary experiment in 2D display (C) and VR setting (D).}
\label{fig:results1}
\end{figure*}

Figure \ref{fig:results1}.A and \ref{fig:results1}.B illustrates the boxplot for the number of successful commands in each trial across 10 trials for VR and 2D screen. Figures \ref{fig:results1}.C and \ref{fig:results1}.D shows the boxplots for the number of false positives in each trail across 10 trails. The boxplots are shown separately for three conditions. First condition is performing pull, push or rotate using only mental command. Second and third conditions are performing the same task using eyebrow movement and smile, respectively. In these figures, the accuracy for a binary experiment is reported. That is, the command is either correctly identified or not. In general, we found that mental commands performed better under VR conditions. The median number of success rate is 8 for VR setting while for the 2D setting the median is 4. However, were still far inferior to facial expressions for practical application usage. In VR, 10 our 10 commands are identified correctly using eyebrow movement in all trails. The median for 2D screen setting is 9 in this case. 

\begin{figure*}
\centering
\includegraphics[width=1\textwidth]{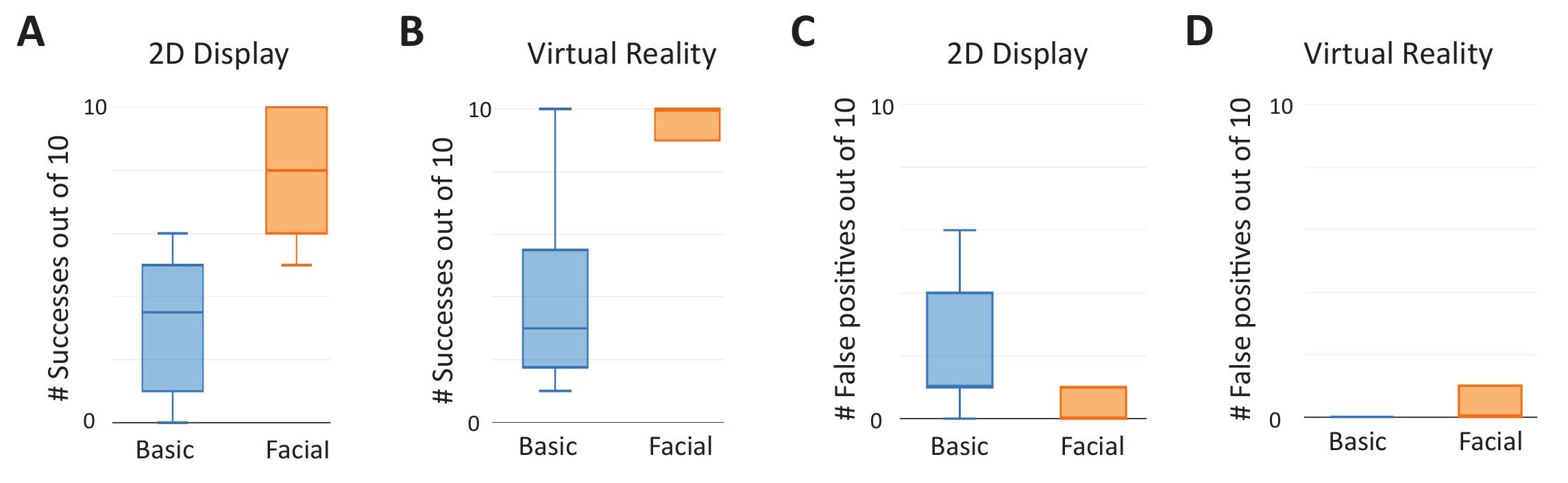}
\caption{Number of successes for trinary experiment in 2D display (A) and VR setting (B). Number of false positives for trinary experiment in 2D display (C) and VR setting (D).}
\label{fig:results2}
\end{figure*}

Figure \ref{fig:results2} shows similar boxplots for the trinary experiments. In this setting, the accuracy is defined is the number of successful identification of a task from three states. The states are object push, object rotation and neutral state. The accuracies are lower compared to binary experiments. Our trials demonstrate that in the 3D environment, mental commands are suitable for binary cases. However, when more than one command is desired, accuracy plummets. In contrast, facial expressions maintain accuracy over all cases.

We found that mental commands were often inconsistent, where different trials would result in drastically different results - perhaps due to movement of the EEG headset. Mental commands were also persistent - they tended to stay active far past when the user wanted the command to stop. In contrast, facial expressions tended to be both consistent and bursty. For comfort, facial expressions were a bit difficult for our subjects while wearing the VR headset, due to the positioning on the face. However, most users still found facial expressions quite usable. Users also reported facial expressions to be more repeatable than mental commands - thinking a command multiple times the same way was found to be rather difficult. In contrast, users quickly picked up facial expressions to accurately and consistently achieve tasks.

We also postulate that facial expressions are much more desirable than mental commands for the wider VR environment, as facial expressions are easily transferable between applications. Meanwhile, mental commands differ greatly between applications, as different applications have different use cases. Due to the length of training, this process can take significant time and effort. Thus, we conclude that for accuracy, a major requirement for most practical VR applications, facial expressions should be used. However, if a more simple "state of mind" is more desirable, mental commands are suitable.
 
\subsection{Carving Brush - VR Application}

\begin{figure}
\centering
\includegraphics[width=0.42\textwidth]{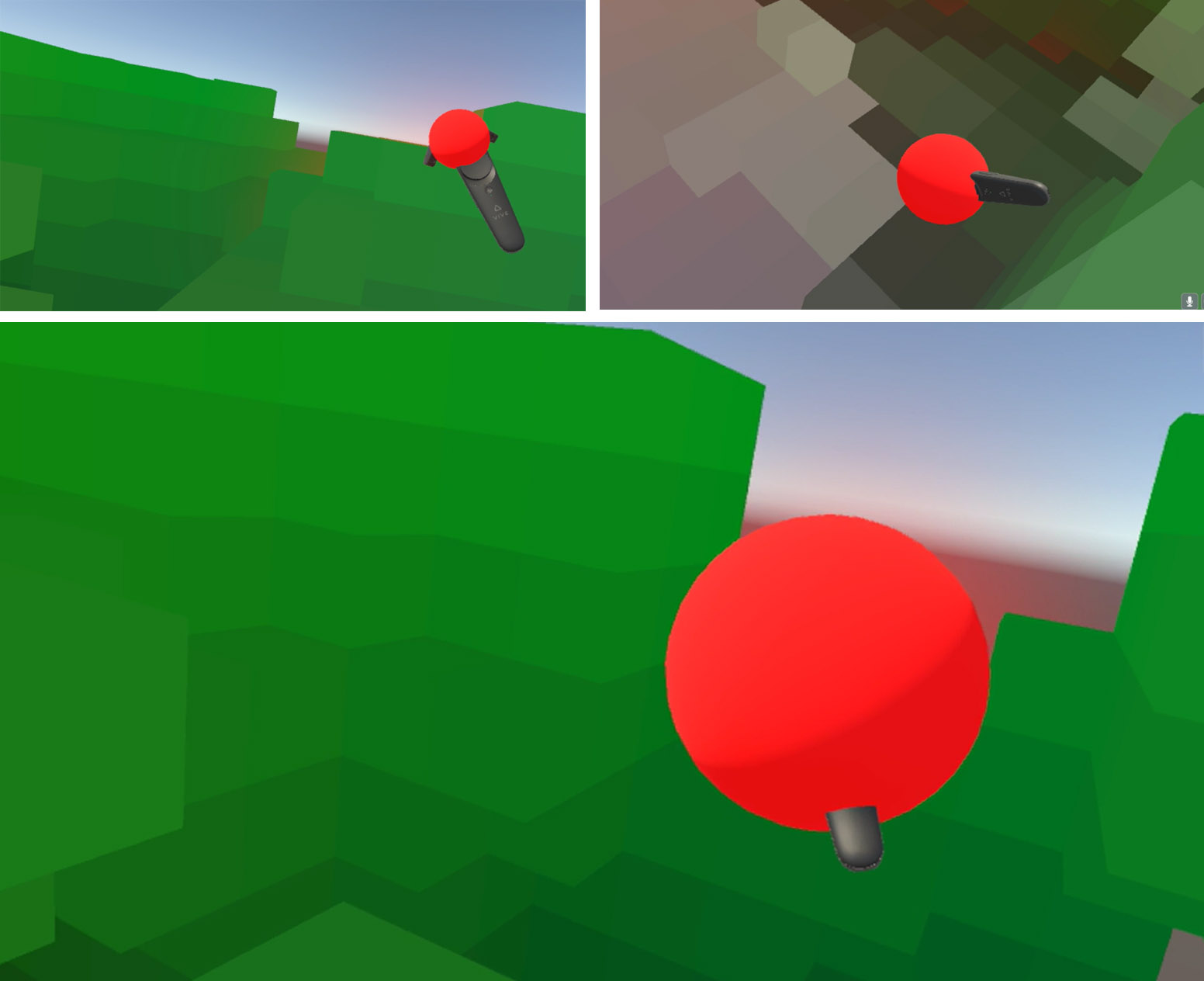}
\caption{VR Carving Brush application}
\label{fig:vr_ap}
\end{figure}

Considering the performance results from each type of command mentioned  above, a VR application is designed to enhance a simple creative process with taking advantage of mental and facial controls. Using a 3D game engine – in our case Unity3D- an array of boxes is generated based on the user’s preference of dimension and scale. These boxes surrounded the user and were rendered with distinguishable colors to form a gradient box. Figure \ref{fig:vr_ap} illustrates three snapshots of this application.

Using the VR controllers as a “destructor brush”, the user can carve out negative volumes from it’s surrounding to make stylized forms and objects with a negative carving technique. This process was done by applying a sphere collider at the top point of the VR controller and deleting each pixel (box) when a collision was detected. In order to optimize the computation process and avoid rendering all the pixel arrays in the initial gradient box, only the outer layer of the object was rendered. With each collision between the brush and specific pixel in space, its surrounding pixels would instantiate to form the updated carved form of the object. The size of the destructor brush changes with mental and facial commands. When the user thinks of “a bigger brush” or raises its eyebrows, the diameter of the sphere collider increases resulting a larger brush for the carving process.  The brush size also decreases when the user thinks of a smaller brush or furrows it brows. The strength of the brush is also designed as a function of speed, which depending on how fast the stroke is made in space, it would change the collider threshold for pixel destruction.

In addition to the size of the brush, changing the color of pixels that surround the brush is also done by mental and facial commands. By thinking of a lighter pixel color or smiling, the surrounding pixels increment their RGB values based on their proximity and position to the VR controller. The closer the controller is to a specific pixel, the more increase of RGB values happen in each frame until they reach the limit of RGB (255, 255, 255) which indicates the color of white.

As this is a 6DOF experience, the user can walk in the space to perform and perform the carving process and mental/facial commands simultaneously . For larger models, a touchpad walking function is implemented which the user navigates the virtual space using touchpad buttons of the VR controller. This method is not recommended as many users’ experience VR sickness (nausea) in such locomotion functions.

\section{Discussion, Assessment and Future Work}

In this paper, we developed a pipeline to assess the application of BCI in VR environment. Our results showed that BCI is more accurate in VR compared to 2D screen. However, ultimately, mental commands as a valid input mechanism for VR applications may need to wait for better technology. After experimenting and analyzing mental and facial commands using EEG sensors in both environments – 2D screens and virtual environments – we believe that due to the inaccurate and noisy inputs, training and implementing commands with current learning methods were not efficient. We believe advanced machine learning methods such as interpretable deep learning tools \cite{Abbasi2017Structural, Abbasi2017Interpreting} and non-linear model estimation algorithms \cite{Abbasi2011estimation} are helpful in enhancing classification problems. 

Also, applying such commands for time-related functions or accurate actions in games and applications can not be applicable at this point. This concern amplifies as we see the number of false negatives increase when a combination of commands is extracted for various tasks. Actions such as eye-movement, unwanted facial expressions, and walking also create noise signals in the EEG recordings. These distortions are unavoidable in some cases to control in the user experience process.

Our future work includes improving the quality of the interface module. Designing an algorithm that is robust to noise in processing the EEG signals is the most important principle in this direction. Developing new applications in virtual-reality such as in the artistic painting, city design, and educational programs are a few examples that require such precision. Robustness to this brain noise also extends beyond EEG devices, so work in this area will certainly help future BCI interfaces that deal with the same issues as EEG devices. We also plan to increase the number of human subjects and tasks in future experiments. This will allow for a more reliable evaluation of our pipeline.

\section*{ACKNOWLEDGMENT}
The authors would like to thank Allen Yang and James F. O’Brien for their constructive feedback.

\end{document}